\documentclass[twocolumn,trackchanges]{aastex701}
\usepackage{graphicx}
\usepackage{amssymb}
\usepackage[fleqn]{amsmath}
\usepackage{verbatim}
\usepackage{natbib} 
\usepackage[utf8]{inputenc}
\usepackage{CJK}
\usepackage{xcolor} 
\usepackage{xspace}

\begin{document}
\defcitealias{wold25}{W25}
\defcitealias{wold22}{W22}
\defcitealias{bouwens21}{B21}

\newcommand{\oii}{\ensuremath{[\text{O~{\sc ii}}]}\xspace}
\newcommand{\nii}{\ensuremath{[\text{N~{\sc ii}}]}}
\newcommand{\oiii}{\ensuremath{[\text{O~{\sc iii}}]}\xspace}
\newcommand{\oiiinosp}{\ensuremath{[\text{O~{\sc iii}}]}}
\newcommand{\hb}{\ensuremath{\text{H}\beta}\xspace}
\newcommand{\oiiihb}{\ensuremath{[\text{O~{\sc iii}}]+\text{H}\beta}\xspace}
\newcommand{\oiiihbnosp}{\ensuremath{[\text{O~{\sc iii}}]+\text{H}\beta}}
\newcommand{\neiiinosp}{\ensuremath{[\text{Ne~{\sc iii}}]}}
\title{Strong \oiiihb emitters dominated the ionizing budget at $z\sim7$}

\author[0000-0002-0784-1852]{Isak G. B. Wold}
\affil{Astrophysics Science Division, Goddard Space Flight Center, Greenbelt, MD 20771, USA}
\affil{Department of Physics, The Catholic University of America, Washington, DC 20064, USA }
\affil{Center for Research and Exploration in Space Science and Technology, NASA/GSFC, Greenbelt, MD 20771}
\email{isak.g.wold@nasa.gov}
\author[0000-0002-9226-5350]{Sangeeta Malhotra} 
\affiliation{Astrophysics Science Division, Goddard Space Flight Center, Greenbelt, MD 20771, USA}
\email{sangeeta.malhotra@nasa.gov}
\author[0000-0002-1501-454X]{James E. Rhoads}
\affiliation{Astrophysics Science Division, Goddard Space Flight Center, Greenbelt, MD 20771, USA}
\email{james.e.rhoads@nasa.gov}
\begin{abstract}

We quantify the ionizing photon production at $z\sim7$ using the deepest spectroscopically confirmed sample of strong \oiiihb emitters (rest-frame EW$>740$\AA) in the Abell 2744 field. Leveraging ultra-deep UNCOVER F410M imaging ($5\sigma\sim29$ AB) and gravitational lensing, we probe an order of magnitude deeper than previous JWST WFSS \oiii\ studies, reaching a luminosity limit of $\log(L_{\oiiihb}/\text{erg s}^{-1})=41.3$. Our rest-frame optical emission-line selection probes some of the youngest, metal- and dust-poor galaxies, identifying a large population of continuum-faint, ionizing candidates. NIRSpec follow-up of a luminosity-representative subset confirms $72\%$ of targets, providing detailed characterization of 18 emitters. Balmer decrements reveal negligible dust, while strong-line diagnostics indicate extremely low metallicities ($12+\log(\text{O/H})=6.8\text{--}7.7$). With typical [O\,\textsc{ii}]/[O\,\textsc{iii}] ratios of $0.054\pm0.007$, we infer an average Lyman continuum escape fraction near the canonical $f_{\text{esc}}=20\%$. Correcting for the spectroscopic confirmation rate, we find that these high-EW emitters represent $56\pm12\%$ of the total UV-selected population by number density. Integrated to our
survey limits, the ionizing budget of these emitters ($\log(\dot{N}_{\rm ion}/{\rm s}^{-1}\,\text{Mpc}^{-3})=50.63\pm0.05$) accounts for $\sim70\%$ of the total budget required for reionization at $z\sim7$. This result is consistent with empirical benchmarks. These results establish \oiiihb selection as a powerful, dust-insensitive probe, showing that known galaxy populations significantly power reionization.

\end{abstract}

\section{Introduction}

Reionization is the last major phase transition of the Universe, when the Intergalactic Medium (IGM) transitioned from neutral to ionized hydrogen. It is thought to be initiated  by the formation of the first stars and galaxies that ionized bubbles around themselves; these bubbles grow and overlap eventually ionizing the Universe. While future $21\text{cm}$ radio observations aim to overcome significant foregrounds to directly map this transition, current observational constraints rely on two bookend signatures: the Gunn-Peterson trough in high-redshift quasar spectra, which suggests reionization was largely complete by $z \sim 6$ \citep[e.g.,][]{fan06}, and the optical depth to Thomson scattering ($\tau$) experienced by Cosmic Microwave Background (CMB) photons, which limits the onset and duration of the epoch.

Our understanding of reionization has shifted significantly with evolving CMB measurements. Early results from WMAP indicated a relatively high optical depth \citep[$\tau = 0.088\pm0.015$;][]{komatsu11}, suggesting an early onset of reionization ($z_\text{mid}\sim11$) that astronomers struggled to explain using known galaxy populations \citep[e.g.,][]{Robertson13}. With more recent Planck results indicating a small optical depth \citep[$\tau=0.054\pm0.007$;][]{planck18} and a later reionization midpoint ($z_\text{mid}\sim7.7$), it was quickly realized that with reasonable assumptions, galaxy populations could explain the needed ionization production \citep[e.g.,][]{robertson15}. While recent studies continue to refine \citep[e.g.,][]{li25} and even challenge \citep[e.g.][]{sailer26,jhaveri25} these CMB constraints, the consensus view has maintained that galaxy populations can supply the required light for reionization.

However, recent JWST observations are complicating this picture. Galaxies at high redshift appear to produce ionizing light more efficiently than previously assumed — quantified by the ionizing production efficiency, $\xi_{ion}$ \citep[e.g.,][]{atek24,simmonds24a}. This leads to a potential over-production of ionizing light, and a revaluation of the reasonable assumptions made about galaxy populations when calculating the ionizing budget \citep[e.g.][]{munoz24,simmonds24b}.

The ionizing photon budget, $\dot{N}_{ion}$, is determined by three variables: the number density of sources (typically derived from UV Luminosity Functions, UVLFs), their ionizing efficiency ($\xi_{ion}$), and the fraction of ionizing light that escapes the host galaxy's interstellar medium ($f_{esc}$).  The determination of both the LF and $\xi_{ion}$ often have to overcome UV-based measurements that are highly sensitive to dust corrections. While, $f_{esc}$ cannot be directly measured due to the opacity of the IGM at high redshifts.

In this work, we explore an emission-line alternative to traditional UV-selected surveys by focusing on the strong \oiiihb population. This selection is significantly less sensitive to dust and allows for the detection of sources regardless of their UV brightness, thereby reducing the reliance on large extrapolations toward the UV-faint end. Following on surveys such as JWST/FRESCO \citep{meyer24}, which pioneered the use of \oiii luminosity functions to constrain the ionizing budget, we advance this approach by reaching approximately an order of magnitude deeper in luminosity. This is achieved by leveraging both the strong gravitational lensing of the Abell 2744 cluster and the ultra-deep UNCOVER F410M imaging ($5\sigma \sim 29$ AB).

Utilizing these data, we calculate $\dot{N}_{ion}$ at $z \sim 7$ using the observed number densities of these emitters combined with constraints from follow-up JWST/NIRSpec G395M/F290LP spectra.  We determine the ionizing budget through two complementary approaches: a direct integration of the \oiiihb LF and an integration of the emitters' UV LF. The first method provides a dust-resilient alternative to UV-based constraints that is independent of $\xi_{ion}$ measurements, while the second method allows us to verify our findings using a standard technique, placing them in the context of the many UV LF studies exploring the ionizing photon budget crisis.

Overall, we find that \oiiihb emitters are ubiquitous at $z\sim7$ and provide the bulk of the ionizing budget required to reionize the universe. We show that extreme solutions to the photon budget crisis — such as a universal $f_{\rm{esc}} < 5\%$ or a sharp turnover in the UV LF — are not required for our $z \sim 7$ sample. We demonstrate that the ionizing photon production rate from \oiiihb emitters remains consistent with empirical constraints down to our survey depth and extrapolate to show that an ionizing photon overproduction crisis is avoided down to depths an order of magnitude deeper ($L_{\oiiihb} \sim 10^{40.5}$ erg s$^{-1}$ or $M_{\text{UV}} \sim -14$).

Throughout this work, all EWs are
rest-frame, \oiii refers to the \oiiinosp$_{5008}$ line while \oiiihb refers to the \oiiinosp$_{5008,4960}$ and H$\beta$ lines.  All magnitudes are in the AB magnitude system ($m_{\mbox{\footnotesize{AB}}}=31.4-2.5$log$_{10}f_{\nu}$
with $f_{\nu}$ in units of nJy). Distances in Mpc units and volumes in Mpc$^3$ units are comoving. We adopt a flat $\Lambda$CDM cosmology
with $\Omega_{m}=0.3$, $\Omega_{\Lambda}=0.7$, and H$_{0}=70$ km
s$^{-1}$Mpc$^{-1}$.

\begin{figure}
\begin{centering}
\includegraphics[width=8.5cm]{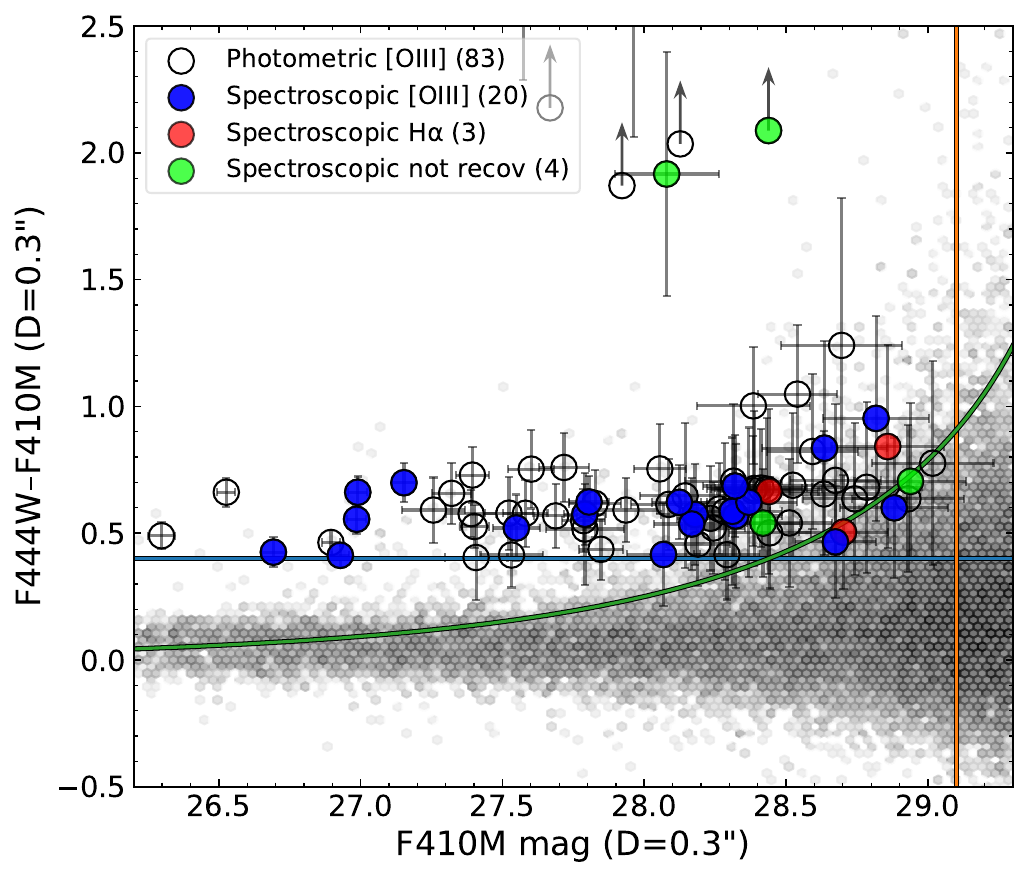}
\caption{F410M-excess selection of $z=7$ \oiiihb emitters in the UNCOVER Abell 2744 field. The overall source distribution is shown by a log$_{10}$-scaled 2D grey density histogram with photometrically selected $z\sim7$ \oiiihb candidates shown by open circles.  Blue, red and green filled circles indicate NIRSpec targets spectroscopically found to be \oiiinosp, H$\alpha$, or non-detections, respectively.    The vertical orange line indicates the median $4.5\sigma$ depth for the UNCOVER MB image.  The horizontal blue line indicates the $0.4$ MB-excess cut used to select EW $>740$\AA\ \oiiihb candidates.  The green curve shows the median $2\Sigma$ cut \citepalias[for details see][]{wold25}. }\label{uncover_confirm}
\par\end{centering}
\end{figure}

\begin{deluxetable}{lccccc}
\tabletypesize{\footnotesize}
\tablecaption{Spectroscopic [OIII] Sample, with Lensing Magnification ($\mu$) \label{tab:sample}}
\tablehead{
\colhead{ID} & \colhead{Obs\#} & \colhead{RA} & \colhead{DEC} & \colhead{Redshift} & \colhead{$\mu$} \\
\colhead{} & \colhead{} & \colhead{(deg)} & \colhead{(deg)} & \colhead{} & \colhead{}
}
\startdata
41004 & 2 & 3.583934 & -30.362109 & 6.769 & 3.35 \\
41006 & 2 & 3.597825 & -30.395942 & 6.872 & 3.59 \\
41007$^{a}$ & 2 & 3.558767 & -30.354825 & 6.998 & 2.25 \\
41008 & 1 & 3.626317 & -30.381901 & 7.070 & 1.47 \\
41009 & 2 & 3.557702 & -30.363984 & 7.228 & 3.28 \\
41018 & 2 & 3.589537 & -30.353730 & 6.772 & 2.19 \\
41026 & 2 & 3.598099 & -30.382377 & 7.291 & 2.22 \\
41028$^{b}$ & 2\textsuperscript{c} & 3.582960 & -30.395225 & 6.870 & 11.42 \\
41035 & 1 & 3.576952 & -30.378847 & 7.329 & 3.77 \\
41038$^{b}$ & 1\textsuperscript{c}\&2 & 3.579663 & -30.398662 & 6.869 & 14.21 \\
41041 & 1\&2\textsuperscript{c} & 3.562740 & -30.383128 & 6.818 & 4.58 \\
41042 & 2 & 3.590824 & -30.347568 & 7.090 & 1.94 \\
41050 & 2 & 3.592703 & -30.366288 & 6.825 & 2.17 \\
41052 & 1 & 3.576326 & -30.419083 & 6.757 & 1.97 \\
41053 & 2 & 3.600120 & -30.365818 & 6.772 & 1.77 \\
41054 & 1\&2 & 3.576688 & -30.383901 & 7.081 & 4.29 \\
41070 & 1 & 3.585672 & -30.374584 & 7.377 & 2.78 \\
41071$^{b}$ & 2 & 3.595798 & -30.393347 & 6.869 & 4.93 \\
41077$^{a}$ & 2 & 3.583480 & -30.350591 & 7.261 & 2.89 \\
41079 & 1 & 3.582044 & -30.407364 & 7.331 & 5.94 \\
\enddata
\tablecomments{
\textsuperscript{a} No UNCOVER DR3 F277W+F356W+F444W counterpart.
\textsuperscript{b} Triply lensed galaxy.
\textsuperscript{c} Severe $\gtrsim 40\%$ slit loss.}
\label{tab:basic}
\end{deluxetable}

\section{Observations}\label{obs}

\subsection{NIRCam F410M target selection of \oiiihb emitters}\label{select}

We select strong rest-frame EW $>740$\AA\ \oiiihb emitters at $z=6.72$-$7.59$ within the Abell 2744 cluster field. Our selection follows the procedure described in \citet[hereafter \citetalias{wold25}]{wold25}, updated to utilize the UNCOVER Data Release 3 (DR3) instead of DR1 \citep[for survey details see][]{bezanson22,furtak23,weaver24,suess24}.  Our \oiiihb selection combines an F410M-excess (F444W-F410M$>0.4$) cut with a requirement for a redshifted Ly$\alpha$ break at $(1+z)1216$ \AA.  

A key DR3 improvement is the addition of ultra-deep F070W JWST imaging ($5\sigma$ depth of 29.6 AB) just blueward of the Ly$\alpha$ break for $z=7$ sources -- making it our most effective veto bandpass.  Unlike the existing deep HST imaging ($17.1$ arcmin$^{2}$ overlap), the F070W is co-aligned with F410M,  providing a $29.3$ arcmin$^{2}$ 
overlapping field of view.  This increased area and depth helps to exclude our sample's main foreground contaminant, H$\alpha$ emitters at $z=5$.

DR3 also delivers an improved F410M mosaic. In DR1 the outer $\sim18''$ border suffered from artifacts and was masked, shrinking the survey area from $29$ to $21$ arcmin$^{2}$ \citepalias{wold25}. In DR3 artifacts are significantly reduced, restoring the full $29$ arcmin$^{2}$ F410M survey footprint.  Additionally, we measure the DR3 F410M mosaic to be 0.1 dex deeper than DR1, bringing the median $5\sigma$ depth to 29.0 AB. In total we identify $N=83$ DR3 candidates over $29$ arcmin$^{2}$ versus $N=68$ in DR1 over $21$ arcmin$^{2}$. We emphasize that an even smaller N=33 sample (14 arcmin$^2$, trimmed F410M image with F814W $5\sigma$ depth $>28$ AB)  was used by \citetalias{wold25} to compute the $z=7$ \oiii LF. In Section \ref{dr3lf}, we show that the DR1 and DR3 LF are consistent, albeit with the larger DR3 survey volume helping to fill in the bright-end of the LF, providing more overlap with the constraints provided by JWST WFSS based results \citep{meyer24,meyer25}.  
Finally, we have updated our analysis to use the new Abell 2744 lensing maps \citep{furtak23,price25}, including their updated list of multiply lensed sources.

\begin{figure*}
\begin{centering}
\includegraphics[width=17.5cm]{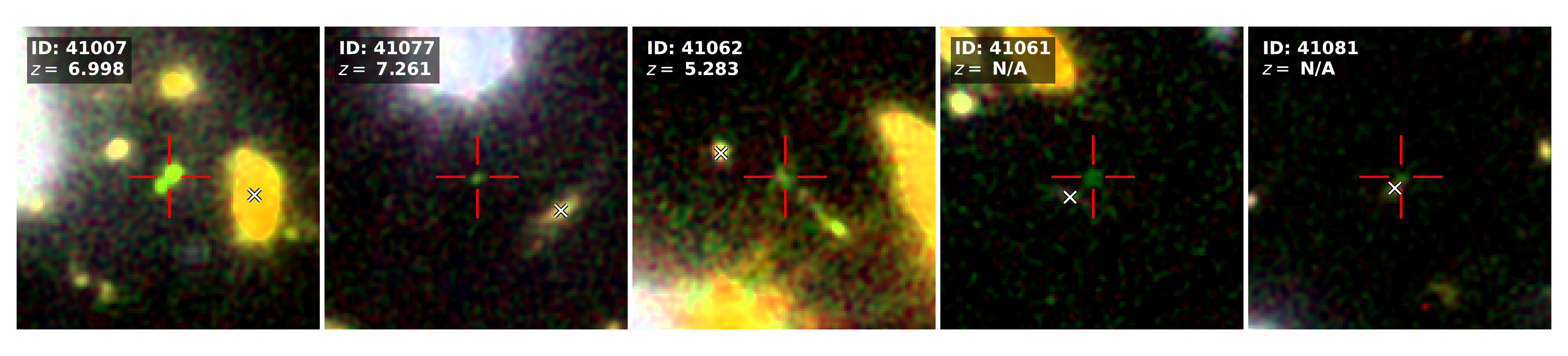}
\caption{$5\times5''$ color images (F200W:blue, F410M:green, F444W:red) of all NIRSpec targeted \oiiihb candidates that have no UNCOVER DR3 counterpart within a $0.1''$ search radius. Two are spectroscopically confirmed to be at $z\sim7$, while one is a $z\sim5$ H$\alpha$ emitter and two have no recovered NIRSpec emission lines.  The small white 'x' indicates the location of the closest DR3 counterpart.  Confirmed objects that have no DR3 counterpart (IDs: 41007, 41077, 41062) are in close proximity to bright stars or galaxies, likely causing DR3 background subtraction issues. }\label{no_dr3_counter}
\par\end{centering}
\end{figure*}

\subsection{NIRSpec G395M Multi-object Spectroscopy}

We followed up a subset of $N=27$ DR3 \oiiihb candidates with NIRSpec G395M/F290LP.  This configuration provides medium resolution ($R=1000$) from 2.87 to 5.10 microns.  We observe rest-frame wavelengths from $3341$ to $6600$ \AA\  covering emission lines including the \oii doublet, \oiiinosp4363, and the Balmer series out to H$\alpha$ for many targets.  We observed these targets with two MSA pointings each with an exposure time of 7.37 hours, designed to detect a 4 micron $10^{-18}$ erg s$^{-1}$ cm$^{-2}$ emission line at a SNR of 15. We used three exposures to nod, taking an exposure in each shutter of our three-shutter slitlet. We did not dither to fill in the NIRSpec chip gap, instead we designed our MSA configurations to prevent the F410M bandpass, containing \oiiinosp5008, from falling within the gap.  The MSA configuration pointing center was optimized using the `midpoint' source centering constraint to keep slit losses $\lesssim 38\%$.  Once this pointing center was determined, we fixed its location and relaxed the centering constraint to encompass the entire open shutter area. This adjustment allowed us to target three additional \oiiihb candidates, albeit at the expense of elevated slit losses ($38\%$ to $70\%$).  Fortunately, these three sources were observed without significant slit losses in our other MSA configuration, one is a multiply-lensed source with spectra of its other lensed component and two have config 1 and config 2 spectra.  We flag the spectra with large slit losses in Table \ref{tab:basic} and omit them from the subsequent analysis.  Reassuringly, unflagged sources observed across multiple configurations exhibit line fluxes that agree within $1\sigma$, while unflagged multiply lensed components show intrinsic line flux agreement within $1.5\sigma$, helping to demonstrate the accuracy of our line flux measurements.

The JWST pipeline version 1.17.1 was used with the Calibration Reference Data System (CRDS) file jwst\_1322 to reduce the NIRSpec G395M/F290LP data.  The standard pipeline procedures were used to produce phase 3 resampled 2D spectra (s2d).  This includes background subtraction, point source pathloss correction, and flux calibration.  From these pipeline generated 2D spectra, we optimally extract 1D spectra with spatial profile weighting \citep{horne86}.  Our objects of interest are continuum faint, so we isolated the bright spectral \oiii5008 line and fit a 1D Gaussian to its spatial profile.  We found our optimal extraction 1D spectra to be consistent with the pipeline's fixed aperture extraction but mitigates off-trace artifacts and has slightly higher signal-to-noise.

\section{Data Analysis}

\subsection{Confirmation rate}\label{crate}

We spectroscopically confirm 20 out of 27 \oiiihb candidates, or 18 out of 25 unique \oiiihb candidates (accounting for a triplely lensed $z\sim7$ galaxy).  Examining the 7 non-\oiii emitters, three are EW $>1000$\AA\ $z=5$ H$\alpha$ emitters and four have no detected spectra/emission lines. In Figure \ref{uncover_confirm} we show our main \oiiihb candidate selection.  An idealized emission line source with no continuum has an expected F410M-excess of 1.0 magnitude, showing that two of the sources without recovered spectra have unphysical colors within their $1\sigma$ error bars, likely indicating artifacts in the F410M mosaic. 
 The three H$\alpha$ emitters have extremely high $>1000$\AA\ rest-frame EWs computed from their 410M-excess, boosting them above our 0.4 magnitude cut.  These H$\alpha$ emitters are very faint (F410M $>28.4$ mag) suggesting that photometric scatter may play an important role in their selection. We confirm $100\%$ of our \oiiihb candidates that have a realistic F410M-excess (F444W-F410M $<1.0$) and are brighter than F410M$=28.4$ mag, corresponding to an \oiiihb luminosity of $\sim10^{41.7}\mu^{-1}$ erg s$^{-1}$, where the $\mu^{-1}$ factor accounts for gravitational magnification.

\begin{figure*}
\begin{centering}
\includegraphics[width=17.5cm]{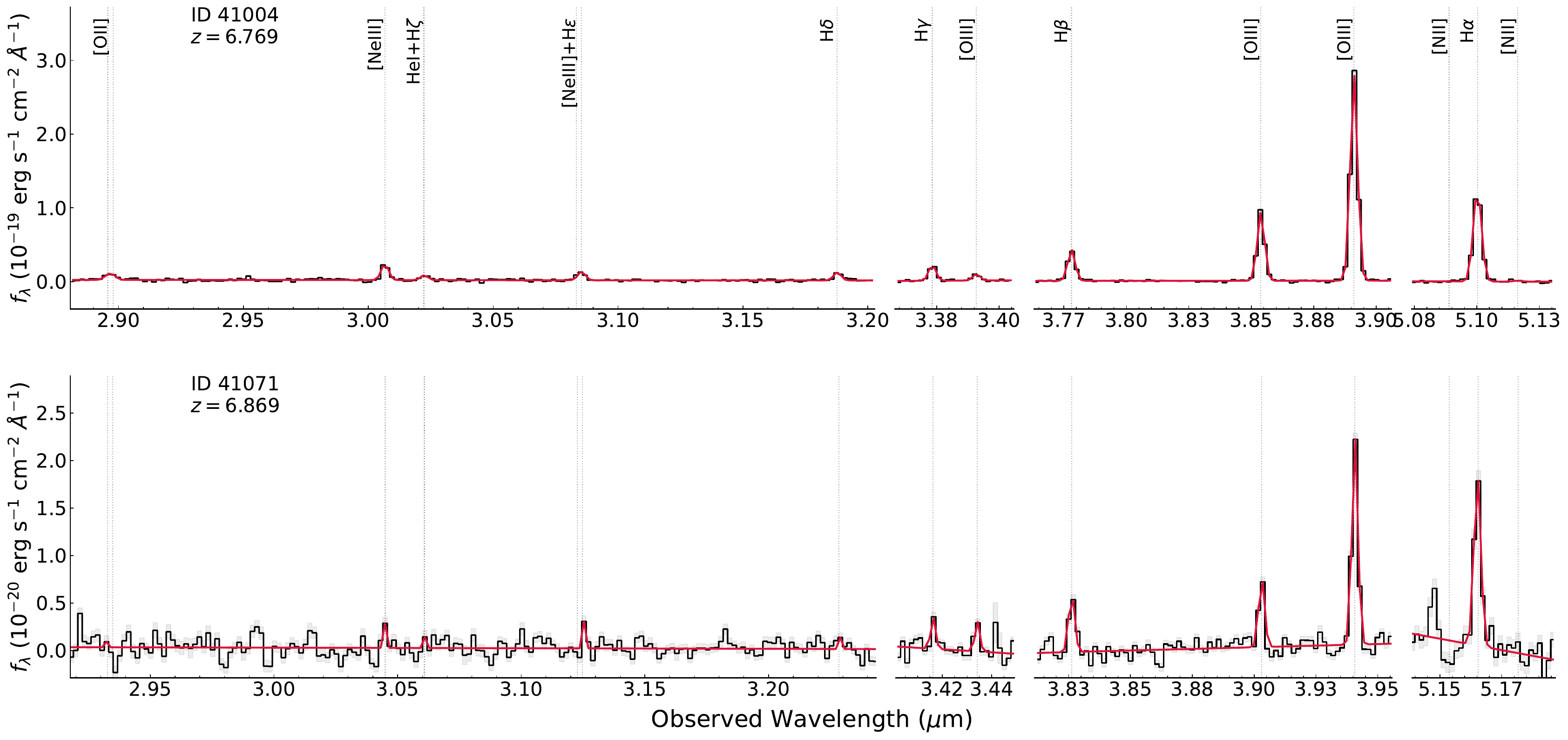}
\caption{Brightest (top panel) and faintest (bottom panel) $z\sim7$ \oiii emitter with full \oii to H$\alpha$ wavelength coverage, demonstrating the quality of our data.  Best-fit models are shown in red.  See Table \ref{tab:sfluxes} for measured line flux values used in this paper.}\label{linefit}
\par\end{centering}
\end{figure*}

\begin{deluxetable*}{lc ccccccc}
\tablecaption{Emission line fluxes in $10^{-18}$ erg s$^{-1}$ cm$^{-2}$ units.\label{tab:fluxes}}
\tablehead{
\colhead{ID} & \colhead{Obs} & \colhead{H$\alpha$} & \colhead{[O III] $\lambda5008$} & 
\colhead{[O III] $\lambda4960$} & \colhead{H$\beta$} & \colhead{H$\gamma$} & 
\colhead{[Ne III] $\lambda3870$} & \colhead{[O II] $\lambda\lambda3727,3730$}
}
\startdata
41004 & 2 & $\mathbf{{ 5.12 \pm 0.13 }}$ & $\mathbf{{ 10.20 \pm 0.18 }}$ & $\mathbf{{ 3.42 \pm 0.06 }}$ & $\mathbf{{ 1.55 \pm 0.07 }}$ & $\mathbf{{ 0.76 \pm 0.07 }}$ & $\mathbf{{ 0.80 \pm 0.09 }}$ & $\mathbf{{ 0.42 \pm 0.12 }}$ \\
41006 & 2 & $\mathbf{{ 2.76 \pm 0.20 }}$ & $\mathbf{{ 3.27 \pm 0.08 }}$ & $\mathbf{{ 1.09 \pm 0.03 }}$ & $\mathbf{{ 1.03 \pm 0.04 }}$ & $\mathbf{{ 0.54 \pm 0.07 }}$ & $\mathbf{{ 0.21 \pm 0.04 }}$ & $0.14 \pm 0.07$ \\
41007\tablenotemark{a} & 2 & \nodata & $\mathbf{{ 3.25 \pm 0.08 }}$ & $\mathbf{{ 1.09 \pm 0.03 }}$ & $\mathbf{{ 0.50 \pm 0.03 }}$ & $\mathbf{{ 0.22 \pm 0.05 }}$ & $\mathbf{{ 0.15 \pm 0.04 }}$ & $\mathbf{{ 0.28 \pm 0.07 }}$ \\
41008 & 1 & \nodata & $\mathbf{{ 3.22 \pm 0.08 }}$ & $\mathbf{{ 1.08 \pm 0.03 }}$ & $\mathbf{{ 0.59 \pm 0.03 }}$ & $\mathbf{{ 0.23 \pm 0.04 }}$ & $\mathbf{{ 0.26 \pm 0.05 }}$ & $0.13 \pm 0.06$ \\
41009 & 2 & \nodata & $\mathbf{{ 2.40 \pm 0.08 }}$ & $\mathbf{{ 0.80 \pm 0.03 }}$ & $\mathbf{{ 0.63 \pm 0.04 }}$ & $\mathbf{{ 0.34 \pm 0.05 }}$ & $\mathbf{{ 0.19 \pm 0.04 }}$ & $0.14 \pm 0.05$ \\
41018 & 2 & $\mathbf{{ 1.41 \pm 0.10 }}$ & $\mathbf{{ 3.64 \pm 0.09 }}$ & $\mathbf{{ 1.22 \pm 0.03 }}$ & $\mathbf{{ 0.54 \pm 0.04 }}$ & \nodata & $\mathbf{{ 0.26 \pm 0.06 }}$ & $\mathbf{{ 0.34 \pm 0.08 }}$ \\
41026 & 2 & \nodata & $\mathbf{{ 1.60 \pm 0.06 }}$ & $\mathbf{{ 0.54 \pm 0.02 }}$ & $\mathbf{{ 0.34 \pm 0.03 }}$ & $\mathbf{{ 0.16 \pm 0.04 }}$ & $0.09 \pm 0.03$ & $0.11 \pm 0.05$ \\
41028\tablenotemark{b,c} & 2 & $\mathbf{{ 2.69 \pm 0.15 }}$ & $\mathbf{{ 2.96 \pm 0.10 }}$ & $\mathbf{{ 0.99 \pm 0.03 }}$ & $\mathbf{{ 0.86 \pm 0.05 }}$ & \nodata & $0.28 \pm 0.09$ & $0.04 \pm 0.11$ \\
41035 & 1 & \nodata & $\mathbf{{ 0.73 \pm 0.10 }}$ & $\mathbf{{ 0.24 \pm 0.03 }}$ & $\mathbf{{ 0.26 \pm 0.06 }}$ & $0.11 \pm 0.10$ & $0.00 \pm 0.04$ & $0.09 \pm 0.07$ \\
41038\tablenotemark{b,c} & 1 & $\mathbf{{ 1.59 \pm 0.14 }}$ & $\mathbf{{ 1.82 \pm 0.09 }}$ & $\mathbf{{ 0.61 \pm 0.03 }}$ & $\mathbf{{ 0.48 \pm 0.05 }}$ & $\mathbf{{ 0.25 \pm 0.06 }}$ & $0.19 \pm 0.07$ & $0.00 \pm 0.09$ \\
41038\tablenotemark{b} & 2 & $\mathbf{{ 1.21 \pm 0.08 }}$ & $\mathbf{{ 1.39 \pm 0.05 }}$ & $\mathbf{{ 0.47 \pm 0.02 }}$ & $\mathbf{{ 0.47 \pm 0.03 }}$ & \nodata & $\mathbf{{ 0.13 \pm 0.04 }}$ & $0.00 \pm 0.05$ \\
41041 & 1 & $\mathbf{{ 0.75 \pm 0.08 }}$ & $\mathbf{{ 0.93 \pm 0.06 }}$ & $\mathbf{{ 0.31 \pm 0.02 }}$ & $\mathbf{{ 0.25 \pm 0.03 }}$ & $\mathbf{{ 0.10 \pm 0.03 }}$ & $0.03 \pm 0.03$ & $0.04 \pm 0.06$ \\
41041\tablenotemark{c} & 2 & $\mathbf{{ 1.22 \pm 0.23 }}$ & $\mathbf{{ 1.60 \pm 0.13 }}$ & $\mathbf{{ 0.54 \pm 0.04 }}$ & $\mathbf{{ 0.44 \pm 0.07 }}$ & $0.22 \pm 0.14$ & $0.22 \pm 0.10$ & $0.11 \pm 0.15$ \\
41042 & 2 & \nodata & $\mathbf{{ 0.71 \pm 0.04 }}$ & $\mathbf{{ 0.24 \pm 0.01 }}$ & $\mathbf{{ 0.28 \pm 0.03 }}$ & $0.08 \pm 0.03$ & $0.00 \pm 0.02$ & $0.02 \pm 0.03$ \\
41050 & 2 & $\mathbf{{ 1.19 \pm 0.14 }}$ & $\mathbf{{ 1.85 \pm 0.10 }}$ & $\mathbf{{ 0.62 \pm 0.03 }}$ & $\mathbf{{ 0.45 \pm 0.05 }}$ & $0.14 \pm 0.06$ & \nodata & \nodata \\
41052 & 1 & $\mathbf{{ 1.25 \pm 0.12 }}$ & $\mathbf{{ 1.13 \pm 0.07 }}$ & $\mathbf{{ 0.38 \pm 0.02 }}$ & $\mathbf{{ 0.38 \pm 0.04 }}$ & $0.15 \pm 0.06$ & \nodata & \nodata \\
41053 & 2 & $\mathbf{{ 0.44 \pm 0.10 }}$ & $\mathbf{{ 1.24 \pm 0.08 }}$ & $\mathbf{{ 0.41 \pm 0.03 }}$ & $\mathbf{{ 0.21 \pm 0.04 }}$ & $0.07 \pm 0.08$ & $0.10 \pm 0.07$ & $0.21 \pm 0.13$ \\
41054 & 1 & \nodata & $\mathbf{{ 0.69 \pm 0.06 }}$ & $\mathbf{{ 0.23 \pm 0.02 }}$ & $\mathbf{{ 0.15 \pm 0.03 }}$ & $0.11 \pm 0.06$ & $0.07 \pm 0.05$ & $0.08 \pm 0.08$ \\
41054 & 2 & \nodata & $\mathbf{{ 0.63 \pm 0.06 }}$ & $\mathbf{{ 0.21 \pm 0.02 }}$ & $\mathbf{{ 0.11 \pm 0.03 }}$ & $0.02 \pm 0.03$ & $0.05 \pm 0.03$ & $0.11 \pm 0.07$ \\
41070 & 1 & \nodata & $\mathbf{{ 0.41 \pm 0.04 }}$ & $\mathbf{{ 0.14 \pm 0.01 }}$ & $\mathbf{{ 0.18 \pm 0.03 }}$ & $0.06 \pm 0.05$ & $0.05 \pm 0.03$ & $0.00 \pm 0.04$ \\
41071\tablenotemark{b} & 2 & $\mathbf{{ 0.63 \pm 0.12 }}$ & $\mathbf{{ 0.66 \pm 0.04 }}$ & $\mathbf{{ 0.22 \pm 0.01 }}$ & $\mathbf{{ 0.19 \pm 0.03 }}$ & $0.08 \pm 0.03$ & $0.06 \pm 0.04$ & $0.01 \pm 0.04$ \\
41077\tablenotemark{a} & 2 & \nodata & $\mathbf{{ 0.48 \pm 0.06 }}$ & $\mathbf{{ 0.16 \pm 0.02 }}$ & $\mathbf{{ 0.18 \pm 0.03 }}$ & \nodata & $0.05 \pm 0.04$ & $0.07 \pm 0.05$ \\
41079 & 1 & \nodata & $\mathbf{{ 0.39 \pm 0.06 }}$ & $\mathbf{{ 0.13 \pm 0.02 }}$ & $\mathbf{{ 0.18 \pm 0.04 }}$ & $0.11 \pm 0.08$ & $0.00 \pm 0.03$ & $0.03 \pm 0.07$ \\
\enddata
\tablecomments{Lines detected above $3\sigma$ are in bold. Fluxes are not corrected for gravitational lensing. Notes a,b,c as defined in Table \ref{tab:sample}.}
\label{tab:sfluxes}
\end{deluxetable*}

 In Section \ref{dr3lf}, we use these spectroscopic results to correct our \oiiihb luminosity function by: 1) removing $N=9$ candidates with unphysical F410M-excess (F410M-excess $1\sigma$ error requires F444W-F410M $>1.0$) and 2) applying a confirmation rate correction ($61.5\%$) for faint emitters (L$_{\oiiihb}<10^{41.7}$ erg s$^{-1}$).  Consistent with \citetalias{wold25}, we also exclude sources that have $<5\%$ recovery rates from our LF sample. These rates are determined via completeness simulations, in which we inject realistic synthetic sources into the science images and evaluate our ability to recover them using the same selection criteria applied to our primary sample. This $<5\%$ completeness cut removes $N=8$ sources from our LF sample including the triply lensed galaxy, counted as three candidates in this tally. The final LF sample has a total of $N=66$ sources.
 
As discussed in \citetalias{wold25}, $20\%$ of our selected candidates lack cataloged counterparts in the F277W+F356W+F444W broadband selected UNCOVER survey \citep{weaver24}. With spectra in hand, we can further investigate this issue. Out of our 25 targeted \oiiihb candidates, five are not found in the DR3 UNCOVER catalog within a $0.1''$ search radius, as shown in Figure \ref{no_dr3_counter}. We spectroscopically confirm two of them as $z\sim7$ \oiiihb emitters, one as a $z\sim5$ H$\alpha$ emitter, and two have no detected spectra/emission lines.  Confirmed objects that have no DR3 counterpart are not our faintest sources, and instead appear to be excluded due to their proximity to bright sources indicating that UNCOVER’s background subtraction procedure inadvertently removed some real sources \citepalias[as suggested in][]{wold25}.

\subsection{Rest-frame optical emission line measurements}

We measured the fluxes for detected emission lines in each spectrum. The NIRSpec MSA confirgurations were designed to always observe the \oiiinosp4960, 5008, and H$\beta$ line triplet.  Other rest-frame optical line fluxes were measured when within the G395M/F290LP bandpass and there was no conflict with a chip gap. To obtain robust fitting results on both strong and
weaker lines, we performed simultaneous gaussian fits of nearby line sets, so that the wavelength and line width were effectively fixed by well-detected lines, allowing more confident extraction of the weak lines \citep[following a similar procedure to][]{rhoads23}. We used the \textsc{SciPy} package \textsc{curve fit} \citep{virtanen20} to obtain best-fit line flux and error measurements using the non-linear least squares method.  We show two examples of fitted spectra in Figure \ref{linefit} which illustrate one of our brighest and one of our faintest emitters.  All line flux measurements used in this paper are listed in Table \ref{tab:sfluxes}.

\section{Results}
\subsection{Spectroscopic Measurements}

Given our sample's \oiiihb luminosity function, we aim to constrain the production rate of ionizing photons.  To accomplish this, we quantify the dependence of the \oiiinosp/H$\beta$ and  \oii/\oiii line ratios on the \oiiihb luminosity.  We also investigate the impact of dust extinction within our sample, as any significant reddening could introduce systematics into our luminosity and ratio measurements if not properly accounted for.  Additionally, we address one object that requires non-standard line fitting due to a H$\alpha$ broad-line component.

\begin{figure}
\begin{centering}
\includegraphics[width=8.5cm]{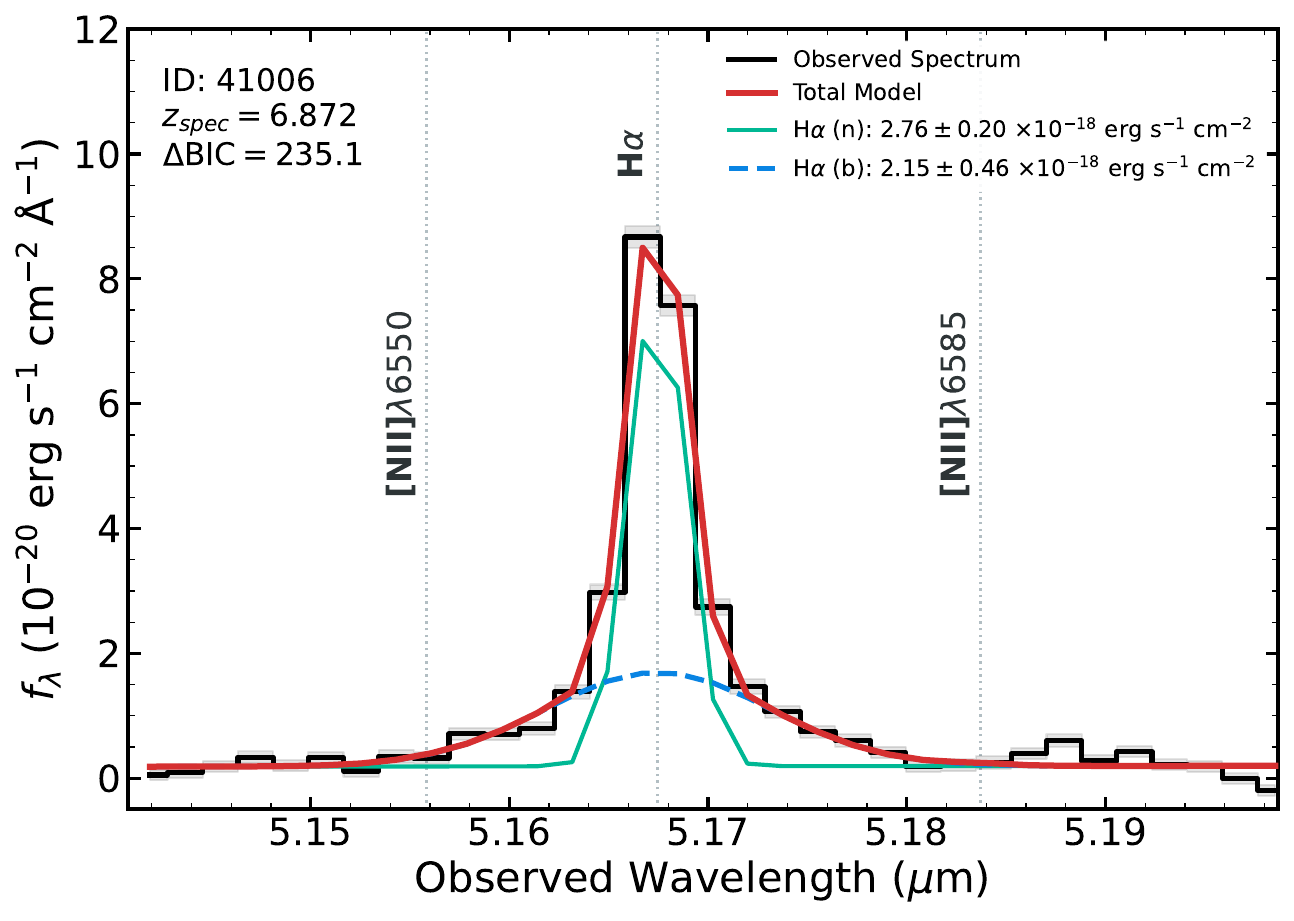}
\caption{Best-fit components for the $H\alpha$ and $\nii\lambda\lambda6550,6585$ complex for object ID 41006. The observed JWST/NIRSpec spectrum is shown in black, the $1\sigma$ error envelope shaded in light gray, and the best-fit model in red.  The inclusion of the broad component is statistically favored, yielding a $\Delta\text{BIC} = 235.1$ compared to a narrow-line-only fit.}\label{lrd}
\par\end{centering}
\end{figure}

\subsubsection{Object ID 41006; Broad H$\alpha$}

We find one object (ID 41006) that has a broad-H$\alpha$ component.  While a detailed analysis of this object is beyond the scope of this work, we find that it is necessary to fit the broad line component to accurately measure the narrow line H$\alpha$ line flux.  Specifically, the broad and narrow components likely originate from different physical scales and undergo varying degrees of dust extinction, a topic we address for the full spectroscopic sample in Section \ref{dust}.

Figure \ref{lrd} displays the best-fit model for the H$\alpha$ spectral region.   Compared to a narrow-line-only fit, the Bayesian Information Criterion \citep[BIC; e.g.,][]{Liddle07} decreases by $\Delta\text{BIC} = 235.1$, showing strong evidence for the additional two broad-line parameters (line width and amplitude). Conversely, we find no significant improvement to the H$\beta$ fit when including a broad component, likely due to the limited signal-to-noise ratio in the $\sim 3\times$ fainter line and/or greater dust attenuation of the broad-line relative to the narrow-line emission \citep[e.g.,][]{veilleux97}. Furthermore, allowing the broad and narrow components to have independent line centers yields no significant improvement to the H$\alpha$ fit.

Following the scaling relations for Little Red Dots (LRDs) from \citet{matthee24}, our best-fit broad-line parameters ($\text{FWHM} = 751 \pm 100 \text{ km s}^{-1}$; $\log_{10}(L_{\text{H}\alpha}/\text{erg s}^{-1}) = 41.5 \pm 0.1$) imply a black hole mass of $\log_{10}(M_{\text{BH}}/M_{\odot}) = 6.1 \pm 0.3$, a bolometric luminosity of $\log_{10}(L_{\text{bol}}/\text{erg s}^{-1}) = 44.0 \pm 0.1$, and an Eddington ratio of $\lambda_{\text{Edd}} = 0.5 \pm 0.3$. For comparison, the LRDs presented in \citet{matthee24} typically exhibit $\log_{10}(M_{\text{BH}}/M_{\odot}) \approx 7\text{--}8$, $\log_{10}(L_{\text{bol}}/\text{erg s}^{-1}) \approx 44.7 \text{--}45.7$, and $\lambda_{\text{Edd}} \approx 0.1\text{--}0.4$. We note the caveat that this source falls below the selection criteria employed by \citeauthor{matthee24} ($L_{\text{H}\alpha} > 2 \times 10^{42} \text{ erg s}^{-1}$, $\text{FWHM} > 1000 \text{ km s}^{-1}$), limits designed to help exclude broad components driven by star-formation-powered outflows. Despite falling below these thresholds, ID 41006 represents an intriguing candidate for a low-mass extension of the LRD population. 

For the present work, this broad H$\alpha$ source plays a very limited role; we simply use the narrow-line H$\alpha$ flux (labeled H$\alpha$ (n) in Figure \ref{lrd}) to determine the Balmer decrement (Section \ref{dust}). Reassuringly, the resulting narrow-line $\text{H}\alpha/\text{H}\beta$ ratio yields a dust attenuation estimate consistent with the $\text{H}\gamma/\text{H}\beta$ decrement, which is measured by our standard fitting procedure and is not affected by broad-line complications.

\subsubsection{Dust extinction}\label{dust}
\begin{figure}
\begin{centering}
\includegraphics[width=8.5cm]{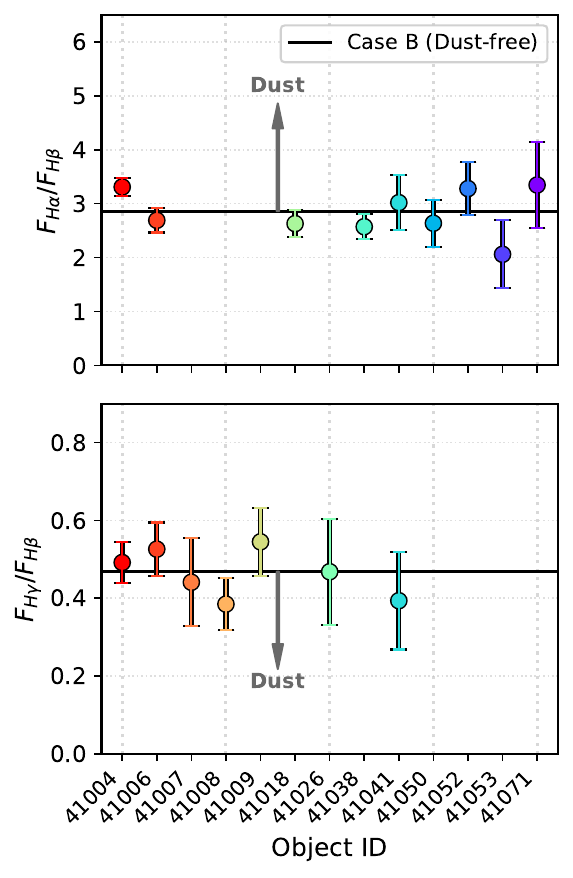}
\caption{Balmer line ratios for spectroscopically targeted \oiiihb emitters. Only lines detected at $>3\sigma$ are shown, with all ratios normalized to H$\beta$.  Observed ratios are consistent with dust-free Case B recombination (horizontal lines) within $3\sigma$. Gray arrows indicate the expected direction of displacement due to dust extinction.}\label{balmer}
\par\end{centering}
\end{figure}

Figure \ref{balmer} shows the Balmer decrement for the \oiiihb emitter spectroscopic sample. All included sources have at least one Balmer line detection ($H\alpha$ or $H\gamma$) at $>3\sigma$ in addition to $H\beta$. We find the sample to be consistent with theoretical dust-free Case B expectations ($T=10,000$ K, $n_e=100$ cm$^{-3}$) within $3\sigma$.

Only one line ratio is found to be $>2\sigma$ from Case B.  This is our brightest \oiii emitter (ID 41004) which is found to have an H$\alpha$/H$\beta$ ratio of $3.31\pm0.16$ compared to the $2.86$ Case B expectation. Assuming a \citet{calzetti00} attenuation law, this implies a dust extinction of A$_{V}=0.6\pm0.3$ mag.  However, the $H\gamma/H\beta$ ratio for this source ($0.49\pm0.05$) is consistent with the dust-free Case B value of $0.47$, yielding an unphysical extinction of A$_{V}=-0.4$ mag with a large $1\sigma$ uncertainty of $\pm0.9$ mag.  A weighted average of both 41004 Balmer ratios results in an extinction of $A_V = 0.5 \pm 0.3$ mag.

Because the Balmer decrement suggests modest extinction for the brightest source and negligible extinction for the remainder of the sample, we proceed without dust extinction corrections. This approach aligns with recent FRESCO results at $z \sim 7-9$, where luminous \oiii emitters ($L_{\oiii} \gtrsim 10^{42}$erg~s$^{-1}$) --- comparable or brighter than our brightest sources --- were found to have effectively zero dust attenuation \citep{meyer24}.

\begin{figure*}
\begin{centering}
\includegraphics[height=7.cm]{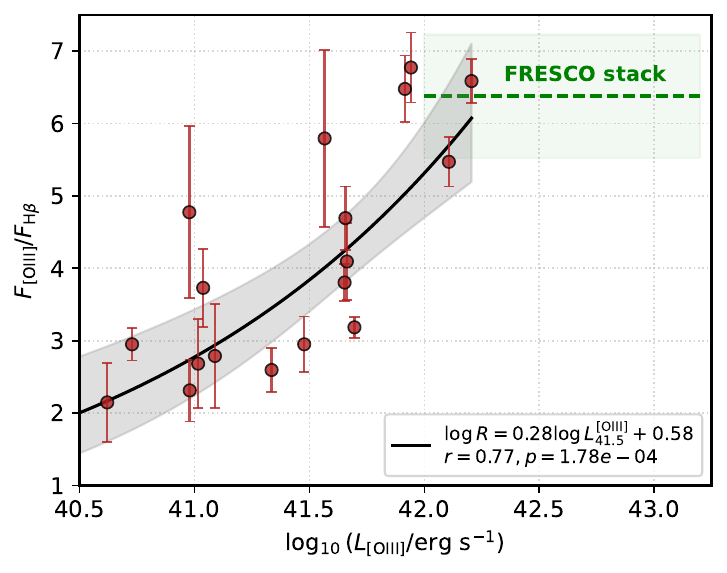}\includegraphics[height=7.cm]{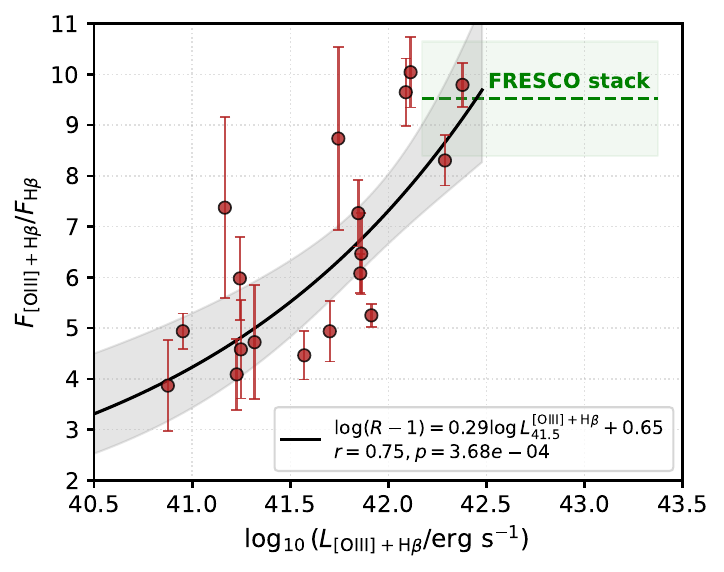}
\caption{\textbf{Left:} \oiiinosp$\lambda$5008 to H$\beta$ line ratio as a function of \oiiinosp$\lambda$5008 luminosity  The luminosity is centered at $10^{41.5} \text{ erg s}^{-1}$ ($L_{41.5}^{\oiii} = L_{[\mathrm{OIII}]}/10^{41.5}$).  The shaded region represents the 95\% confidence interval of the fit. The FRESCO stack line ratio of $6.38 \pm 0.85$ \citep[green dashed line;][]{meyer24} is consistent with our most luminous sources. \textbf{Right:} Same as the left panel, but for the \oiiihb triplet. The power-law functional form is adopted to enforce the physical constraint that $(\oiiihb)/\mathrm{H}\beta \geq 1$ as $L_{\oiii} \to 0$. This fit is utilized in Section \ref{ion} to derive ionizing photon production from the \oiiihb luminosity function. In the legend of both panels, $R$ indicates the y-axis ratio.}\label{o3hb}
\par\end{centering}
\end{figure*}

\subsubsection{\oiiihb line ratio}\label{o3hb_sec}

We investigate the relationship between the \oiiinosp$\lambda5008/\mathrm{H}\beta$ line ratio and \oiii luminosity to help constrain the ionizing production of the sample. As shown in the left panel of Figure \ref{o3hb}, we observe a trend where the \oiiinosp$\lambda5008/\mathrm{H}\beta$ ratio increases with \oiii luminosity. While our most luminous sources ($\log_{10} L_{[\mathrm{OIII}]} \gtrsim 42$) are in excellent agreement with the stacked results from FRESCO (\oiiinosp$\lambda5008/\mathrm{H}\beta=6.38 \pm 0.85$; \citealt{meyer24}), our individual measurements reveal that this ratio drops significantly at lower luminosities. 

To ensure that this trend is not a mathematical artifact induced by the shared \oiii variable across the axes, we performed a linear regression directly on the \oiii and \hb normalized log-luminosities, $L_{41.0}^{\mathrm{H}\beta} = L_{\mathrm{H}\beta}/10^{41.0}$ and $L_{41.5}^{\mathrm{[OIII]}} = L_{\mathrm{[OIII]}}/10^{41.5} \text{ erg s}^{-1}$:

\begin{equation}
\log_{10} L_{41.0}^{\mathrm{H}\beta} = (0.72 \pm 0.06) \log_{10} L_{41.5}^{\mathrm{[OIII]}} - (0.08 \pm 0.03).
\end{equation}

\noindent A constant line ratio (slope $m=1$) is rejected at the $5.1\sigma$ level, confirming that applying a single value across the entire luminosity function will misrepresent the ionizing output of the fainter galaxy population (see Section \ref{ion}).

Using the $z = 2\text{--}9$ strong-line calibrations from \citet{sanders24}, our measured \oiiinosp$\lambda5008/\mathrm{H}\beta$ ratios --- which provide the highest signal-to-noise among our available indicators --- yield metallicities ranging from $12+\log(\mathrm{O/H}) = 6.8\text{--}7.7$. These values represent roughly $1\text{--}10\%$ of the solar abundance and are consistent with results derived from local analogs of extreme emission line galaxies \citep{jiang19}. This range assumes our sample occupies the low-metallicity branch of the inherently double-valued \oiiinosp$/\mathrm{H}\beta$ indicator.  To break this degeneracy, we use the single-valued \neiiinosp$\lambda3869/$\oii calibration \citep{sanders24}; given the weighted mean of \neiiinosp$\lambda3869/$\oii$=1.6\pm0.3$, we find a metallicity of $12+\log(\mathrm{O/H}) \sim 7.4$. This result is consistent with the $\oiiinosp/\mathrm{H}\beta$ low-metallicity branch, confirming our initial branch selection.  We defer direct metallicity measurements and a comprehensive investigation of the emission line properties to future work --- incorporating full SED modeling to further refine these physical constraints --- as the present analysis focuses specifically on the properties required to characterize the ionizing budget.

We note that our F410M-selection does not have the resolution to isolate \oiiinosp$\lambda5008$ by itself and instead selects strong \oiiihb emitters.  Given the correlation seen between the \oiiinosp$\lambda5008/\mathrm{H}\beta$ line ratio and \oiii luminosity, in Section \ref{dr3lf} we compute the \oiiihb luminosity function and then consistently account for this dependence when computing the ionizing photon production in Section \ref{ion}.  To accomplish this, in the right panel of Figure \ref{o3hb} we quantify the relationship between the $(\oiiihb)/\mathrm{H}\beta$ ratio and \oiiihb luminosity. The $(\oiiihb)/\mathrm{H}\beta$ ratio is inherently bounded at unity, representing the limit where \oiii emission becomes negligible relative to the $\mathrm{H}\beta$ recombination line. We adopt a functional form that asymptotically approaches this physical floor to prevent unphysical extrapolations at the low-luminosity end of our sample. Thus, we adopt a linear $\log_{10}(R-1)$ vs.\ $\log_{10}L_{\oiiihb}$ functional form,  where $R$ is the $(\oiiihb)/\mathrm{H}\beta$ line ratio, indicating that the ratio minus one scales as a power-law with \oiiihb line luminosity ($R-1 \propto L_{\oiiihb}^{0.29}$).  We use this scaling relation in our computation of the ionizing budget in Section \ref{ion}.

\subsubsection{\oii/\oiii line ratio}

\begin{figure}
\begin{centering}
\includegraphics[width=8.5cm]{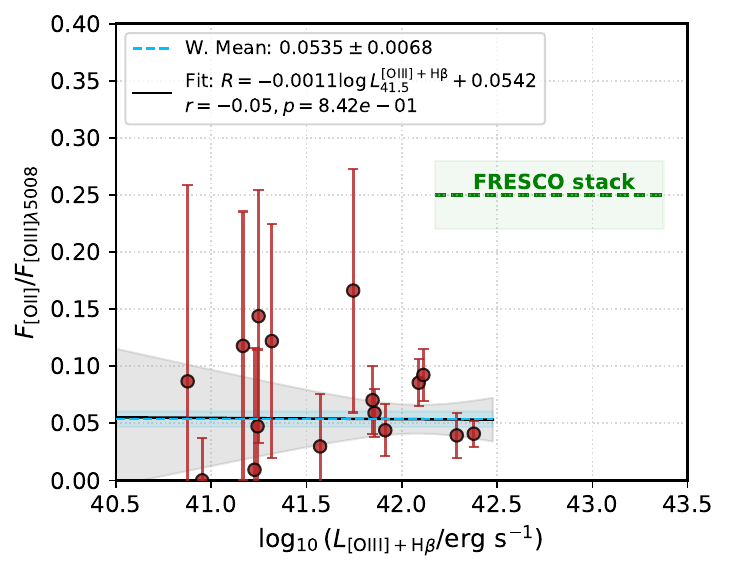}
\caption{Line ratio of \oii$\lambda\lambda$3727,3729 to \oiiinosp$\lambda$5008---a ratio found to correlate with Lyman continuum (LyC) leakage---plotted against the \oiiihb luminosity. We find no significant correlation ($r = -0.04$, $p = 0.88$), suggesting that the physical conditions favoring LyC leakage are independent of \oiiihb luminosity across this range. The gray shaded region represents the 95\% confidence interval of the linear fit, with the sample's weighted mean shown in blue. Notably, our sample exhibits consistently lower ratios than the FRESCO stack \citep[$0.25 \pm 0.02$; green dashed line][]{meyer24}, potentially indicating more favorable conditions for ionizing photon escape or stacking limitations.}\label{o2o3}
\par\end{centering}
\end{figure}

We investigate the ratio of $[\mathrm{OII}]\lambda\lambda3727,3729$ to $[\mathrm{OIII}]\lambda5008$ (hereafter $O_{23}$) as a function of \oiiihb luminosity in Figure \ref{o2o3}. The $O_{23}$ ratio (or more commonly its inverse $O_{32}$) is a widely used proxy for the ionization parameter and has been shown in both local and intermediate-redshift samples to correlate with the escape fraction of Lyman continuum (LyC) photons \citep[e.g.,][]{izotov18, nakajima20,flury22}. High ionization states (low $O_{23}$) typically trace the density-bounded H~II regions favorable for LyC leakage.  We adopt $O_{23}$ over the more traditional $O_{32}$ to ensure numerical stability, as our $[\mathrm{OII}]$ fluxes are extremely faint and the $O_{32}$ ratio becomes highly sensitive to small noise fluctuations in the denominator.

We find no significant correlation between $O_{23}$ and $L_{\oiiihb}$ ($r = -0.04$, $p = 0.88$); the near-zero Pearson coefficient and high $p$-value confirm that a linear relationship is absent. This stands in contrast to the strong luminosity dependence observed for the $\oiii/\mathrm{H}\beta$ ratio (see Section \ref{o3hb_sec}). This suggests that the physical conditions governing the ionization state of the ISM—and by extension, the potential for LyC escape—are not strongly coupled to the \oiiihb line luminosity of these systems. 

Furthermore, our sample exhibits consistently lower $O_{23}$ values compared to the FRESCO stack at $z \sim 7-9$ \citep[$0.25 \pm 0.02$;][]{meyer24}. Our weighted mean of $O_{23}=0.054\pm0.007$ (indicated by the blue line in Figure \ref{o2o3}) suggests significantly higher ionization parameters, potentially indicating more favorable conditions
for ionizing photon escape for our sample. Alternatively, this difference may highlight inherent stacking limitations in the FRESCO study, where the underlying biases and averaged properties of a composite sample fail to capture the characteristic $O_{23}$ ratio. 

Since $f_{esc}$ cannot be measured directly for these sources, we rely on local analogs to estimate its value. Using the $O_{32}$--$f_{esc}$ relations from \citet{chisholm22}, which are based on the Low-redshift Lyman Continuum Survey \citep[LzLCS;][]{flury22}, we find that our sample's weighted mean \oii/\oiii ratio ($O_{23} = 0.054 \pm 0.007$) corresponds to $f_{esc}$ values ranging from $18\%$ to $27\%$. This motivates our adoption of $f_{esc} \sim 20\%$ when evaluating the ionizing production of our sample (see Section \ref{ion}). 

Recent JWST studies have inferred the escape fractions of reionization-era galaxies via spectral energy distribution (SED) fitting, finding no significant dependence of $f_{esc}$ on UV luminosity \citep{papovich26,giovinazzo26}. This is broadly consistent with our own results, in which we find no evidence for an $f_{esc}$ dependence on \oiiihb line luminosity.  These two SED studies report different typical $f_{esc}$ values—one finding $f_{esc}<5\%$ while the other finds an average of $\sim10\%$, which is more in line with our findings. Our higher $f_{esc}\sim20\%$ value is consistent with a high-EW selection; such sources likely trace young, intense starbursts where low-density channels in the ISM allow for the efficient escape of ionizing radiation.

 \begin{figure}
\begin{centering}
\includegraphics[width=8.3cm]{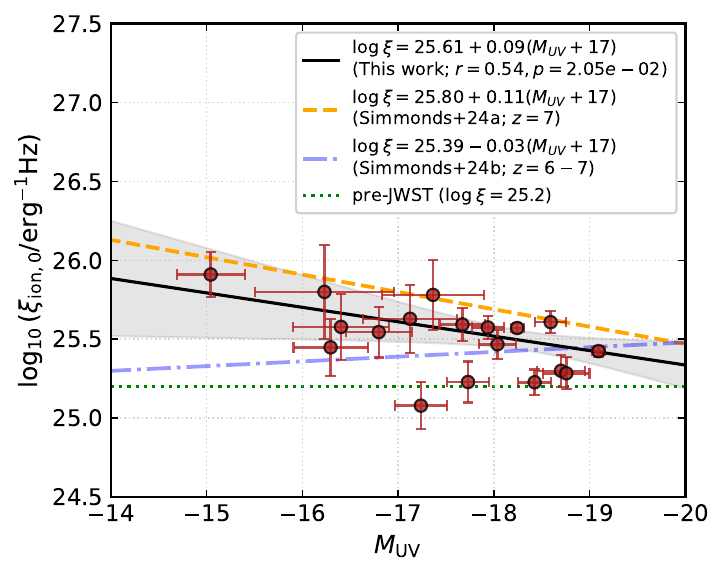}
\caption{Ionizing production efficiency, $\xi_{\rm{ion,0}}$, as a function of absolute UV magnitude ($M_{\rm{UV}}$) at $z \sim 7$. Our spectroscopic sample is shown as red points, with the best-fit relation (solid black line) of: $\log \xi = 25.61 + 0.09(M_{UV} + 17)$. The gray shaded region represents the $95\%$ confidence interval. Our results show an increase in efficiency toward fainter magnitudes, aligning with the emission-line selected sample from \citet[][orange dashed line]{simmonds24a}. In contrast, the mass-complete sample from \citet[][blue dash-dotted line]{simmonds24b} exhibits a lower-efficiency trend.  The horizontal green dotted line indicates the canonical pre-JWST efficiency of $\log_{10}(\xi_{\rm{ion,0}}) = 25.2$ \text{erg}$^{-1}$ \text{Hz} \citep[e.g.,][]{robertson15}.}\label{xi}
\par\end{centering}
\end{figure}

 \begin{figure*}
\begin{centering}
\includegraphics[width=18cm]{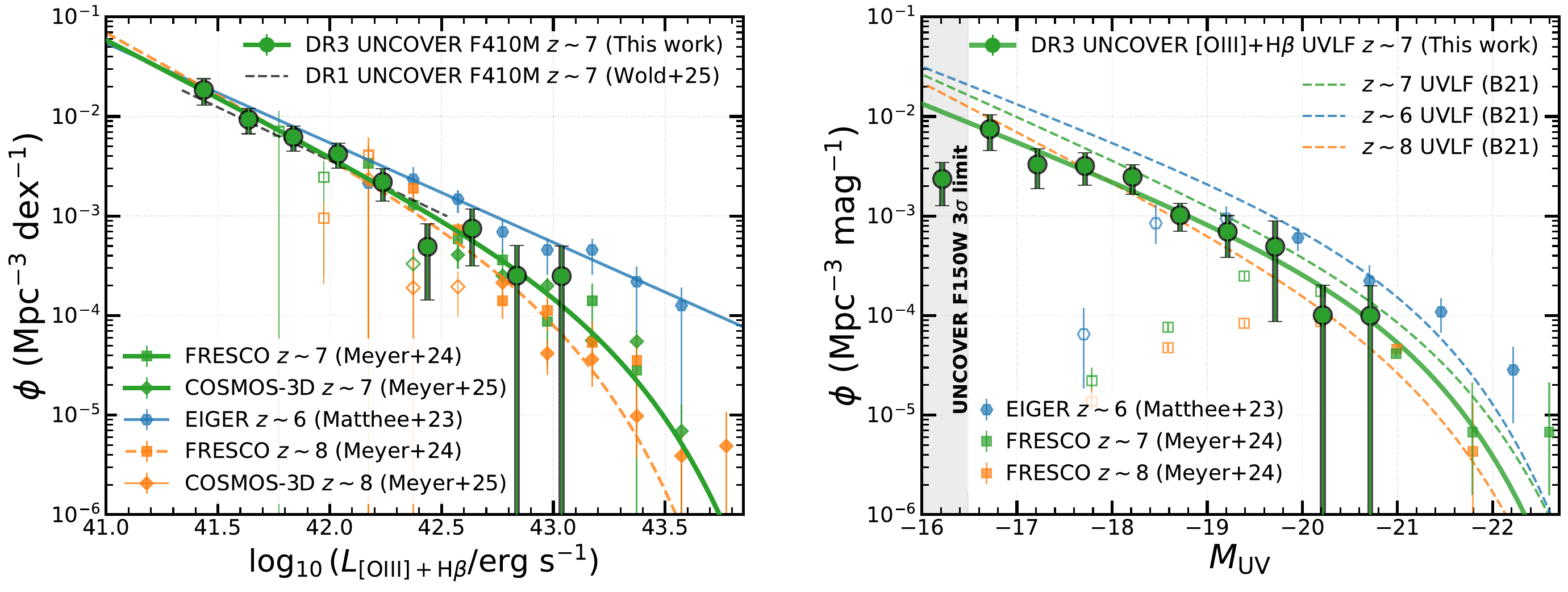}
\caption{\textbf{\textit{Left:}}The DR3 UNCOVER \oiiihb luminosity function compared to LF results from the literature  \citep{matthee23,meyer24,wold25,meyer25}. \textbf{\textit{Right:}} The UV luminosity function of DR3 UNCOVER \oiiihb emitters (green points).  The number densities are comparable to UV-selected galaxies \citep[dashed green curve;][]{bouwens21}, demonstrating that strong \oiiihb emission is common at $z\sim7$.    Line-selected WFSS studies (EIGER and FRESCO) become incomplete at magnitudes fainter than -20, only picking up the extreme upper end of the EW distribution.  At brighter magnitudes the WFSS surveys have number densities comparable to the UVLFs, consistent with our results.}\label{o3Hblf}
\par\end{centering}
\end{figure*}

\subsection{Absolute UV magnitudes and $\xi_{ion}$}\label{uv_meas}

We compute absolute UV magnitudes ($M_{\mathrm{UV}}$) for our sample to: (1) determine the density of these extreme emitters ($\oiiihb > 740$\,\AA) relative to $z=7$ UV-continuum selected sources; and (2) evaluate their cumulative ionizing production based on the UV luminosity function, ensuring consistency with our $[\text{O}\,\textsc{iii}]+\mathrm{H}\beta$ luminosity function results. This second goal requires the computation of the ionization efficiency, $\xi_{\rm{ion}}$, which converts the UV luminosity density to the ionization production rate.

For our $z\sim7$ sample, we determine their absolute UV magnitude $M_{\mathrm{UV}}^{1500\text{\AA}}$ with F150W, assuming a UV slope of $\beta=-2$, where $f_{\lambda} \propto \lambda^{\beta}$.  The F150W imaging reaches a $3\sigma$ depth of $30.05$ mag -- detecting $89\%$ of the LF sample -- while its effective wavelength ensures that UV magnitude measurements remain relatively insensitive to spectral slope variations. Additionally, the F150W band remains free of Lyman-alpha break contamination for all sources, ensuring reliable UV magnitude measurements even for our highest-redshift objects ($7.33 < z < 7.59$) where the break begins to enter the F115W window.

Both the spectroscopic and photometric sample have UV slopes that are consistent with the canonical value of $\beta = -2$ at the $1.5\sigma$ level. Specifically, we find spectroscopically confirmed sources with Lyman alpha break blue-ward of the F115W bandpass ($z<7.33$) and $\sigma_{\beta}<0.5$ has a weighted mean of $\beta=-2.21\pm0.24$, while the larger  photometric sample with $\sigma_{\beta}<0.5$ has a weighted mean of $\beta=-1.85\pm0.10$. We note this photometric result is potentially biased toward redder (more positive) values; without spectroscopic redshifts, this sample likely includes sources where the Lyman-alpha break has begun to enter the F115W bandpass, artificially suppressing the blue-filter flux and inflating the resulting slope. Here the UV slope is determined from the F115W and F150W flux where,

\begin{equation}
    \frac{f_{\nu}^{\mathrm{F150W}}}{f_{\nu}^{\mathrm{F115W}}} = \left( \frac{\lambda_{\mathrm{pivot}}^{\mathrm{F150W}}}{\lambda_{\mathrm{pivot}}^{\mathrm{F115W}}} \right)^{\beta + 2}
,\end{equation}

and

\begin{equation}
\sigma_{\beta} = \frac{\sqrt{ \left( \frac{\sigma_{\nu}^{\mathrm{F150W}}}{f_{\nu}^{\mathrm{F150W}}} \right)^2 + \left( \frac{\sigma_{\nu}^{\mathrm{F115W}}}{f_{\nu}^{\mathrm{F115W}}} \right)^2}}{\ln(\lambda_{\mathrm{pivot}}^{\mathrm{F150W}} / \lambda_{\mathrm{pivot}}^{\mathrm{F115W}})}.
\end{equation}

We compute the ionization efficiency $\xi_{\rm{ion},0} = \frac{L_{H\beta}}{c_{H\beta}L_{\text{UV}}}$, assuming  dust-free, $T_{e} = 10^4$ K, and $n_e = 100$ cm$^{-3}$ Case B recombination with a line-emission coefficient $c_{H\beta} = 4.86 \times 10^{-13}$ erg \citep[e.g.,][]{schaerer03,matthee23}. Non-zero LyC escape 
reduces recombination line production, which
is accounted for via $\xi_{\rm{ion}} = \frac{\xi_{\rm{ion},0}}{1-f_{esc}}$, where $\xi_{\rm{ion},0}$ assumes $f_{esc}=0$.  Figure \ref{xi} shows that $\xi_{\rm{ion},0}$ increases for fainter galaxies within our spectroscopic sample.  This trend is consistent with \citet{simmonds24a} that studied 677 emission line galaxies from $z=4$ to $9$, but at odds with \citet{simmonds24b} which studied a larger mass complete sample ($>10^{7.5}$ M$_\odot$ at the $90\%$ level) over a similar redshift range (orange versus blue line in Figure \ref{xi}).  \citet{simmonds24b} explain this discrepancy as a selection effect, where their original emission line selection preferentially selects high $\xi_{\rm{ion}}$ galaxies.  Given our selection for strong \oiiihb emission, it is expected that our results align more closely with the emission-line selected sample of \citet{simmonds24a} rather than the mass-complete sample of \citet{simmonds24b}.  We apply our sample's best-fit $\xi_{\rm{ion},0}$ dependence on M$_{UV}$ when computing the ionization budget in Section \ref{ion}, limiting its application to our high-EW sample.

\begin{deluxetable*}{lcccccc}
\tabletypesize{\footnotesize}
\tablecaption{Best-fit Schechter Parameters and Integrated Quantities to Survey Depth \\
$\log L > 41.34$, $M_{\rm UV} < -16.49$ \label{tab:schechter}}
\tablehead{
\colhead{$z\sim7$ Sample} &
\colhead{$\log\phi^{\ast}$\tablenotemark{a}} &
\colhead{$\log L^{\ast}$ or $M^{\ast}$\tablenotemark{b}} &
\colhead{$\alpha$} &
\colhead{$\log n$\tablenotemark{c}} &
\colhead{$\log\rho_{L}$\tablenotemark{d}} &
\colhead{$\log\dot{N}_{\rm ion}$\tablenotemark{e}}
}
\startdata
[OIII]+H$\beta$ LF & $-4.12^{+0.42}_{-0.53}$ & $43.19^{+0.31}_{-0.24}$ & $-2.15^{+0.13}_{-0.12}$ & $-2.07^{+0.06}_{-0.06}$ & $39.79^{+0.04}_{-0.04}$ & $50.63^{+0.04}_{-0.05}$ \\
{[OIII]}+H$\beta$ UVLF & $-3.66^{+0.41}_{-0.49}$ & $-20.81^{+0.61}_{-0.72}$ & $-1.95^{+0.15}_{-0.13}$ & $-2.04^{+0.03}_{-0.03}$ & $25.79^{+0.04}_{-0.04}$ & $50.67^{+0.03}_{-0.03}$ \\
\citetalias{bouwens21} UVLF & $-3.72^{+0.15}_{-0.16}$ & $-21.15^{+0.13}_{-0.13}$ & $-2.06^{+0.11}_{-0.11}$ & $-1.79^{+0.08}_{-0.09}$ & $26.01^{+0.05}_{-0.05}$ & \nodata \\
\enddata
\tablenotetext{a}{$\log_{10}(\phi^{\ast}/\mathrm{Mpc}^{-3})$.}
\tablenotetext{b}{$\log_{10}(L^{\ast}/\mathrm{erg\,s}^{-1})$ or $M^{\ast}$ (AB mag).}
\tablenotetext{c}{$\log_{10}(n/\mathrm{Mpc}^{-3})$}
\tablenotetext{d}{$\log_{10}(\rho_L)$ in erg\,s$^{-1}$\,Mpc$^{-3}$ ([OIII]+H$\beta$ LF) or erg\,s$^{-1}$\,Hz$^{-1}$\,Mpc$^{-3}$ (UVLFs).}
\tablenotetext{e}{$\log_{10}(\dot{N}_{\rm ion})$ in photons\,s$^{-1}$\,Mpc$^{-3}$, assuming $f_{\rm esc}=0.2$. }
\end{deluxetable*}
\subsection{DR3 LFs}\label{dr3lf}

We compute the rest-frame EW $>740$\AA\ \oiiihb LF at $z=6.72$-$7.59$ using the procedure described in \citetalias{wold25}, updated from UNCOVER DR1 to DR3 and incorporating the spectroscopic corrections described in Section \ref{crate}.  Figure \ref{dr3lf} shows that the DR3 LF (green circles) is in agreement with the best-fit DR1 LF (dashed black power-law) and with the $z\sim7$ LFs from the literature \citep[green squares and diamonds;][]{meyer24,meyer25}.  The main DR3 LF difference from UNCOVER DR1 is better contraints on the bright-end due to a larger survey volume.  DR1 was limited to regions with deep HST veto band coverage, while DR3 adds a deep F070W veto bandpass over the entire F410M region of interest. This increased bright-end coverage confirms the agreement between $z\sim7$ LFs computed from F410M-excess and slitless-spectroscopic surveys in the $10^{42.2}$ to $10^{42.9}$ erg s$^{-1}$ luminosity range.

Consistent with \citetalias{wold25}, we fit the $z\sim7$ UNCOVER LF with a Schetcher function using the \citeauthor{meyer24}\ FRESCO data to constrain the bright-end.  We use the observed FRESCO \oiiinosp5008,4960+H$\beta$ line ratio of $1.49$ to convert their \oiiinosp5008 LF to a \oiiinosp5008,4960+H$\beta$ LF.  The best-fit parameters are obtained by 
minimizing the Cash statistic \citep{cash79}, which is appropriate for 
Poisson-distributed count data, and uncertainties are derived from 
$10{,}000$ Monte Carlo realizations in which the observed counts in each 
bin are Poisson-resampled and the fit is repeated; the 16th and 84th 
percentiles of the resulting parameter distributions define the $1\sigma$ 
confidence intervals.  The left panel of Figure \ref{o3Hblf} shows the best fit Schechter function for the combined $z\sim7$ LF, with parameters $\alpha = -2.15^{+0.13}_{-0.12}$, $\rm{log}_{10}(\phi^*/Mpc^{-3}) = -4.12^{+0.42}_{-0.53}$, and $\rm{log}_{10}(L^*/\rm{erg\ s}^{-1}) = 43.19^{+0.31}_{-0.24}$, where the Schechter function \citep{schechter76} is defined as:

\begin{equation}
\Phi(L)dL=\phi^{*}\left(\frac{L}{L^{*}}\right)^{\alpha}e^{-L/L^{*}}d\left(\frac{L}{L^{*}}\right)
\end{equation}

Comparison with $z=6$ to $z=8$ LFs from the literature shows tentative evidence for a decline in the LF with increasing redshift.  While there is large variation in the best-fit LFs at $z\sim8$, this is likely caused by highly incomplete data-points.  In Figure \ref{o3Hblf}, FRESCO and COSMOS-3D data points with completeness $<25\%$ are shown as open symbols, demonstrating that the WFSS LFs agree once this completeness threshold is exceeded.

\begin{figure*}
\begin{centering}
\includegraphics[width=18cm]{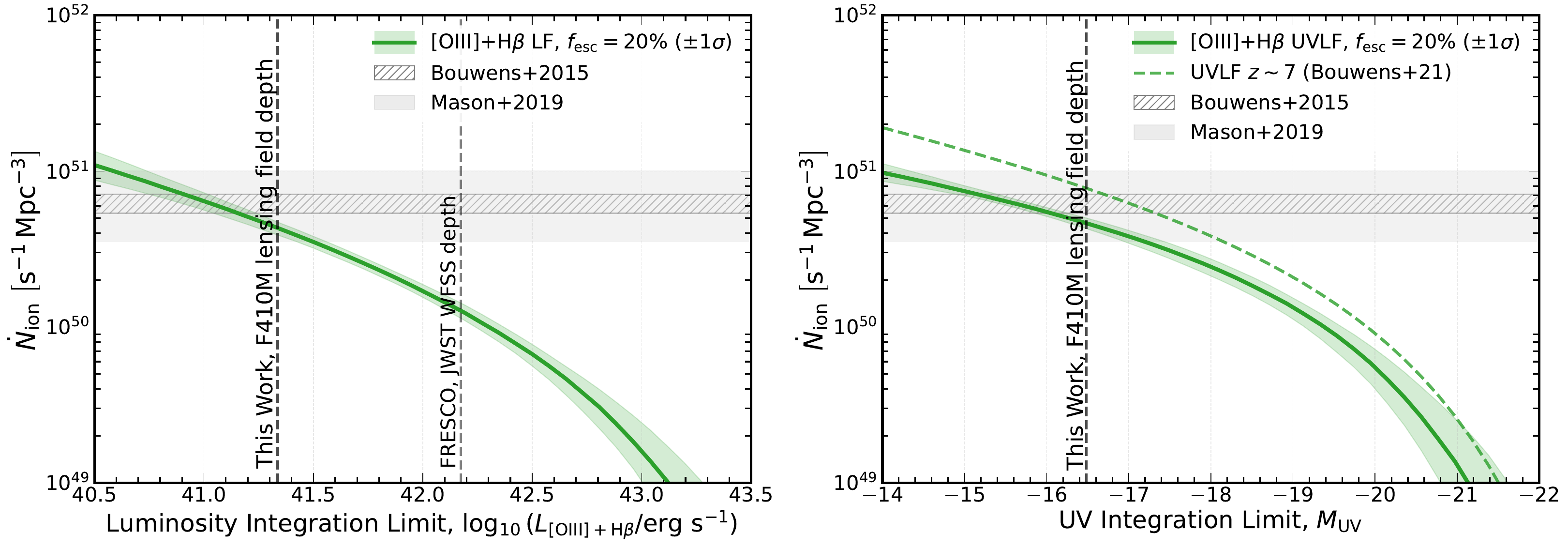}
\caption{The ionizing photon rate density, $\dot{N}_{\text{ion}}$, as a function of survey depth at $z \sim 7$. \textbf{\textit{Left:}} $\dot{N}_{\text{ion}}$ derived from the $[\text{O~III}]+\text{H}\beta$ luminosity function. \textbf{\textit{Right:}} $\dot{N}_{\text{ion}}$ derived from the UV luminosity function of \oiiihb emitters. In both panels, the solid green curve represents the constraints from our high-EW sample, while the dashed green curve (Right) assumes the ionizing properties of high-EW sources apply to the entire galaxy population. Horizontal shaded regions indicate the $z=7$ benchmarks from \citet{bouwens15} and \citet{mason19}, inferred from Planck CMB and IGM neutral fraction constraints. We find consistent results between both methods, indicating that our sample accounts for $\sim 70\%$ of the required ionizing budget using $f_{esc}=20\%$ derived from the mean $O_{23}$ ratio of our spectroscopic sample. Our results are consistent with  $\dot{N}_{\text{ion}}$ constraints until surveys reach depths an order of magnitude beyond our current limits, at which point an extrapolation of our high-EW constraints reaches $>100\%$ of the ionizing budget ($L_{\oiiihb}\sim10^{40.5}$ erg s$^{-1}$  or $M_{UV} \sim -14$) and highlights the potential for a photon over-production crisis. The green shaded region shows the $\pm1\sigma$ uncertainty on $\dot{N}_{\rm ion}$, 
propagated from the MC chains of the Schechter fit parameters 
(Section~\ref{dr3lf}); at each integration depth, we compute 
$\dot{N}_{\rm ion}$ for each MC realization and take the 16th and 84th 
percentiles as the uncertainty bounds.}

\label{o3Hbndot}
\par\end{centering}
\end{figure*}

We also compute the $z\sim7$ UV luminosity function of strong \oiiihb emitters using the sample's absolute UV magnitudes (Section~\ref{uv_meas}). As previously done for the \oiiihb LF fit, we again use the \citeauthor{meyer24}\ data to constrain the bright-end.  The right panel of Figure \ref{o3Hblf} shows the resulting best fit Schechter function with parameters $\alpha = -1.95^{+0.15}_{-0.13}$, $\rm{log}_{10}(\phi^*/Mpc^{-3}) = -3.66^{+0.41}_{-0.49}$, and $M^* = -20.81^{+0.61}_{-0.72}$, where the UV Schechter function is defined as:

\begin{equation}
\Phi(M)dM = \frac{\ln10}{2.5} \phi^{*} \left[10^{0.4(M^{*}-M)}\right]^{\alpha+1} e^{-10^{0.4(M^{*}-M)}} dM
\end{equation}

Using the \citetalias{bouwens21} $z\sim7$ UVLF as a reference, strong \oiiihb emitters are common at $z\sim7$, comprising $56\pm12\%$ of the overall UV-selected population by number density and $60\pm9\%$ by UV luminosity density to our survey depth ($M_{UV}<-16.5$; see Table \ref{tab:schechter} for all computed densities). As an alternative way to constrain the bright-end, fixing $M^{\ast}=-21.15$ and $\alpha=-2.06$ to the 
\citetalias{bouwens21} $z\sim7$ values and fitting only the normalization to our UVLF data
gives $\phi^{\ast}_{\rm fit}/\phi^{\ast}_{\rm B21} = 60\pm5\%$, consistent with the number density comparison above. Assuming the EW distribution is luminosity-independent and follows an exponential form ($\propto e^{-\mathrm{EW}/EW_0}$) this fraction implies a scale length of $EW_0 = 1450$\,\AA\ for \oiiihbnosp, or equivalently $EW_0 \sim 950$\,\AA\ for \oiiinosp5008 alone, which is consistent with  \oiiinosp5008 EW distributions measured in archival IRAC 
and JWST studies \citep{Labbe13, smit15, endsley21, matthee23}.

Referring to the right panel of Figure \ref{o3Hblf}, line-selected WFSS studies (EIGER and FRESCO) become incomplete at UV magnitudes fainter than $\sim -20$, only picking up the extreme upper end of the EW distribution, explaining their UVLF decline at faint magnitudes.   For this reason, we only use their $<-20$ M$_{\rm{UV}}$ data (solid green squares in Figure \ref{o3Hblf}) to constrain the bright-end of our best-fit Schechter function.

Overall the evolution of the \oiiihb LFs and their corresponding UV LFs show evidence for a gradual decline from $z=6$ to $z=8$, albeit with increased scatter at the faint-end, likely driven by highly incomplete WFSS surveys at these low luminosities.  Best-fit Schechter parameters and derived integrated quantities — number density and luminosity density — are listed in Table~\ref{tab:schechter} for both LFs and compared to \citetalias{bouwens21}.

\subsection{Ionizing Budget based on \oiiihb and UV LFs}\label{ion}

We determine the contribution of strong \oiiihb emitters to the $z \sim 7$ ionizing budget, $\dot{N}_{\text{ion}}$, using two complementary approaches: a direct integration of the \oiiihb LF and an integration of the UV LF of \oiiihb emitters. The first method is directly constrained by our F410M-excess selection and provides a robust alternative to UV-based constraints, as it is independent of $\xi_{\text{ion}}$ measurements and less sensitive to dust attenuation. The second method serves to verify these results using a more commonly employed technique and places them in context of the many UV LF studies exploring the ionizing photon budget crisis.  In both cases, the calculation can be viewed as  a conversion of the LF in question to a H$\beta$ luminosity density, which is then used to compute the ionizing production rate required for a given Lyman continuum escape fraction.

Using the \oiiihb LF, $\Phi$, we compute the ionizing budget as:

\begin{equation}
\dot{N}_{\text{ion}} = \int_{L_{\min}}^{\infty}  f_{esc} \, \dot{n}_{\text{ion}} \, \Phi \,   dL_{[\text{OIII}]+H\beta}
\end{equation}

\noindent where all the factors inside the integral can, in principle, depend on $L_{\oiiihb}$. However, as discussed in Section \ref{o2o3}, we assume a constant $f_{esc}\sim20\%$ given the mean $O_{23}$ ratio of our spectroscopic sample.  Here, $\dot{n}_{\text{ion}}$ is the production rate of ionizing photons for an individual galaxy with s$^{-1}$ units:
 
\begin{equation}
\label{eq:o3_ndot}
\dot{n}_{\text{ion}} =  \frac{L_{[\text{OIII}]+H\beta}}{R\,c_{H\beta}(1-f_{esc})}
\end{equation}

\noindent $R$ is the luminosity-dependent line ratio, $([\mathrm{OIII}] + \mathrm{H}\beta)/\mathrm{H}\beta$ (relation shown in the right panel of Figure \ref{o3hb}), and $c_{H\beta}=4.86\times10^{-13}$ erg is the H$\beta$ emission line coefficient, or the energy produced in H$\beta$ emission per recombination event, assuming case B recombination ($T_{e} = 10^4$ K, $n_e = 100$ cm$^{-3}$ e.g., \citealt[][]{schaerer03,matthee23}). The factor $(1-f_{esc})$ in the denominator accounts for the escape of ionizing radiation that does not contribute to nebular emission.  This formulation is consistent with a picket-fence model, in which ionizing radiation escapes through low-column density channels in an otherwise optically thick ISM \citep[e.g.,][]{saldana-lopez22}. For a similar $\dot{N}_{\text{ion}}$ calculation, see \citet{meyer24}.

Using the UV LF of \oiiihb emitters, $\Phi_{UV}$, we also compute the ionizing budget as:

\begin{equation}
\dot{N}_{\text{ion}} = \int_{-\infty}^{M_{\text{lim}}}  f_{esc} \, \dot{n}_{\text{ion}} \, \Phi_{UV} \,   dM_{UV}
\end{equation}

\noindent where $\dot{n}_{\text{ion}}$ is now:

\begin{equation}
\label{eq:uv_ndot}
\dot{n}_{\text{ion}} =  \frac{L_{UV}\, \xi_{\rm{ion},0}}{(1-f_{esc})}
\end{equation}

\noindent The ionization efficiency is determined from our spectroscopic sample via $\xi_{\rm{ion},0} = \frac{L_{H\beta}}{c_{H\beta}L_{\text{UV}}}$, highlighting the similarity of Equations \eqref{eq:o3_ndot} and \eqref{eq:uv_ndot}.  We employ a M$_{UV}$-dependent $\xi_{\rm{ion},0}$ as defined in Section \ref{uv_meas}.

In Figure \ref{o3Hbndot}, we show the ionizing photon budget, $\dot{N}^{}_{\text{ion}}$, as a function of survey depth for both methods and compare our results to the $z=7$ constraints derived by \citet{bouwens15} and \citet{mason19}.  These \citeauthor{bouwens15}\ and \citeauthor{mason19}\ benchmarks represent empirical determinations of the cosmic ionizing emissivity inferred from Planck CMB optical depth measurements - which set the upper bound - and IGM neutral fraction constraints, which provide a lower bound. Consistent with  \citeauthor{meyer24}\footnote{modified to our assumed $f_{esc}=20\%$}, we find that FRESCO's WFSS survey depth can account for $\sim20\%$ of the ionizing output needed for reionization of the Universe at $z\sim7$, while our deeper survey accounts for $\sim70\%$.  Given the range of allowable $\dot{N}_{\text{ion}}$ values, we find that our sample can account for $60$-$85\%$ of the \citeauthor{bouwens15}\ ionizing budget or $45$-$100\%$ of the more lenient \citeauthor{mason19}\ ionizing budget using $f_{esc}=20\%$.  Generalizing to arbitrary $f_{\rm esc}$, 
\begin{equation}
    \dot{N}_{\rm ion} = \dot{N}^{20\%}_{\rm ion} \times 
    \frac{f_{\rm esc}/(1-f_{\rm esc})}{0.25},
\end{equation}
where $\dot{N}^{20\%}_{\rm ion} = 10^{50.63\pm0.05}$\,s$^{-1}$\,Mpc$^{-3}$ is our \oiiihb LF value integrated down to the survey limit
($\log L_{\oiiihb} > 41.34$). The denominator $0.25$ normalizes the relation to our fiducial assumption of $f_{\rm esc}=0.2$. For comparison, consistent estimates of $\dot{N}_{\rm ion}$ derived from the \oiiihb UVLF are provided in Table \ref{tab:schechter}.

In right panel of Figure \ref{o3Hbndot} (dashed green curve), we also show that if we apply the $\xi_{ion,0}$ and $f_{esc}=20\%$ factors derived for our high-EW sample to the full \citetalias{bouwens21} $z\sim7$ UVLF, the entire ionizing budget can be accounted for at a depth of $M_{UV}\sim-17$.  This is the same $M_{UV}$ threshold identified by \citet{munoz24} that is needed to explain $100\%$ of the ionizing output required for reionization; integrating beyond this threshold results in over-production and the proposed photon budget crisis.  However, \citet{simmonds24b} argues that $\xi_{ion,0}$ factors derived from high-EW samples should not be applied to the entire population - the very assumption made by \citeauthor{munoz24}, and represented here by our dashed green curve.  \citeauthor{simmonds24b}\ find that more representative, mass-complete samples exhibit a milder scaling of $\xi_{ion,0}$ with $M_{UV}$, avoiding an over-production of ionizing light down to at least $M_{UV} \sim -16$. This is roughly the same depth to which our high-EW $\dot{N}_{\text{ion}}$ curve (solid green curve) must be extrapolated to account for $100\%$ of the ionizing budget, suggesting that high-EW galaxies are the main reionization drivers since these sources likely represent the highly efficient subset within the broader \citeauthor{simmonds24b}\ population.

Our results emphasize how common high-EW sources are at $z\sim7$, yet our extrapolated $\dot{N}_{\text{ion}}$ constraints do not become inconsistent with empirical benchmarks --- suggesting an over-production of ionizing photons --- until surveys reach an order of magnitude deeper  ($L_{\oiiihb}\sim10^{40.5}$ erg s$^{-1}$ or $M_{UV}\sim-14$), which is beyond even the deepest published UVLFs \citep[e.g., $M_{UV}\sim-15$ to $-16$;][]{finkelstein22,atek24}, although see the recently submitted results by \citet{atek26} for constraints down to $M_{UV}\sim-12$. Overall, a consistent picture is emerging where current samples significantly contribute to the ionizing budget --- driving reionization --- but we argue that deeper surveys with spectroscopic followup are essential to balance the ionizing budget by determining whether the trends observed in our high-EW sample hold at fainter luminosities.

\section{Summary}

In this work, we have utilized the gravitational lensing of Abell 2744 and ultra-deep UNCOVER F410M medium-band imaging to investigate the ionizing budget at $z \sim 7$. Our \oiiihb selection is significantly less affected by dust than standard UV-based surveys and identifies a large population of continuum-faint, highly efficient ionizers. 

Our JWST/NIRSpec follow-up spans the photometric sample's luminosity range and characterizes this high-EW \oiiihb population as having negligible dust extinction, low gas-phase metallicities of $12+\log(\mathrm{O/H}) \approx 6.8\text{--}7.7$ ($1\text{--}10\%$ solar), and an average escape fraction near the canonical $f_{\text{esc}}=20\%$ value.

We find that high-EW \oiiihb emitters are a dominant component of the $z \sim 7$ galaxy population. By comparing the UV LF of our emitters to the total star-forming UV LF, we determine they constitute $ \sim60\%$ of galaxies by number density. Furthermore, their contribution to the ionizing budget is substantial; whether calculated via direct integration of the \oiiihb LF or through the emitters' UV LF, this population provides $\sim 70\%$ of the required ionizing budget at our survey’s depth.

These findings demonstrate that the observed population of emitters can account for a high percentage -- possibly all -- of the ionizing light needed to drive cosmic reionization at $z=7$.  This means that there is little need for more exotic sources to contribute significantly.  Given our survey depth, we do not find evidence for an over-production of ionizing photons - the photon budget crisis - but taking our results and extrapolating to deeper depths this crisis can quickly arise, emphasizing the importance of future deep surveys with spectroscopic followup.

\noindent \begin{acknowledgments} The material is based upon work supported by NASA under award number 80GSFC24M0006. This work is based on observations made with the NASA/ESA/CSA James Webb Space Telescope. The data were obtained from the Mikulski Archive for Space Telescopes at the Space Telescope Science Institute, which is operated by the Association of Universities for Research in Astronomy, Inc., under NASA contract NAS 5-03127 for JWST. These observations are associated with program \#6053. Support for program \#6053 was provided by NASA through a grant from the Space Telescope Science Institute, which is operated by the Association of Universities for Research in Astronomy, Inc., under NASA contract NAS 5-03127. The specific \#6053 observations analyzed can be accessed via \dataset[https://doi.org/10.17909/1czm-c662]{https://doi.org/10.17909/1czm-c662}. This work was supported in part by the
Nancy Grace Roman Space Telescope Wide Field Preparatory
Science Award 22-ROMAN22-0090. The authors acknowledge the UNCOVER team led by PIs Labb\'{e} \& Bezanson, JWST \#2561, for developing their observing program with a zero-exclusive-access period.

\end{acknowledgments}

\bibliography{wold}
\bibliographystyle{aasjournalv7}
\end{document}